\documentclass[prd,twocolumn,superscriptaddress,preprintnumbers,nofootinbib,dvipsnames]{revtex4}
\usepackage{graphicx}
\usepackage{epsfig}
\usepackage{bm}
\usepackage{latexsym,amssymb,amsmath,amsfonts,amssymb,txfonts,pxfonts,wasysym,float}
\usepackage{mathrsfs}
\usepackage{xcolor}
\usepackage{lettrine}
\usepackage{lipsum}
\usepackage{enumitem}
\usepackage{appendix}

\usepackage{hyperref}
\hypersetup{
    colorlinks=true, % false: boxed links; true: colored links
    linkcolor=blue, % color of internal links (change box color with linkbordercolor)
    citecolor=magenta}

\def\gs{\mathrel{
   \rlap{\raise 0.511ex \hbox{$>$}}{\lower 0.511ex \hbox{$\sim$}}}}
\def\ls{\mathrel{
   \rlap{\raise 0.511ex \hbox{$<$}}{\lower 0.511ex \hbox{$\sim$}}}}

\begin{document}

\title{Neutrino constraints on long-lived heavy dark sector particle decays in the Earth}

\author{Mary Hall Reno}
\address{Department of Physics and Astronomy,
University of Iowa, Iowa City, IA 52242, USA}

\author{Luis A. Anchordoqui}
\affiliation{Department of Physics \& Astronomy,  Lehman College, City University of
  New York, NY 10468, USA
}
\affiliation{Department of Physics,
 Graduate Center, City University
  of New York,  NY 10016, USA
}

\affiliation{Department of Astrophysics,
 American Museum of Natural History, NY
 10024, USA
}

\author{Atri Bhattacharya}
\affiliation{Space Sciences, Technologies and Astrophysics Research (STAR) Institute, Universit\'e de Li\`ege, B\^at. B5a, 4000 Li\`ege, Belgium}

\author{Austin Cummings}
\affiliation{ Department of Physics, Pennsylvania State University, State College, Pennsylvania 16801 USA}

\author{Johannes Eser}
\affiliation{Department of Astronomy \& Astrophysics, KICP, The University of Chicago, Chicago, IL 60637, USA}

\author{Claire~Gu\'epin}
\affiliation{University of Maryland, College Park, Maryland}

\author{John~\nolinebreak~F.~\nolinebreak~Krizmanic}
\affiliation{CRESST/NASA Goddard Space Flight Center, Greenbelt, MD 20771, USA \\
University of Maryland, Baltimore County, Baltimore, MD 21250, USA}

\author{Angela~\nolinebreak~V.~\nolinebreak~Olinto}

\affiliation{Department of Astronomy \& Astrophysics, KICP, The University of Chicago, Chicago, IL 60637, USA}

\author{Thomas Paul}
\affiliation{Department of Physics \& Astronomy,  Lehman College, City University of
  New York, NY 10468, USA
}

\author{Ina Sarcevic}
\affiliation{Department of Physics, University of Arizona, Tucson, Arizona 85721, USA}
\affiliation{Department of Astronomy and Steward Observatory,
University of Arizona, Tucson, Arizona 85721, USA}

\author{Tonia M. Venters}
\affiliation{Astrophysics Science Division, NASA Goddard Space Flight Center, Greenbelt, MD 20771, USA}

\date{\today} 

\begin{abstract}
  \noindent 
  Recent theoretical work has explored dark matter accumulation in the Earth and its drift towards the center of the Earth that, for the current age of the Earth, does not necessarily result in a concentration of dark matter ($\chi$) in the Earth's core. We consider a scenario of long-lived ($\tau_\chi\sim 10^{28}$ s), super heavy ($m_\chi=10^7-10^{10}$ GeV) dark matter that decays via $\chi\to \nu_\tau \bar{\nu}_\tau$ or $\chi\to \nu_\mu \bar{\nu}_\mu$. We show that an IceCube-like detector over 10 years can constrain a dark matter density that mirrors the Earth's density or has a uniform density with density fraction $\epsilon_\rho$ combined with the partial decay width $B_{\chi\to \nu_\tau \bar{\nu}_\tau}\Gamma_\chi$ in the range of $(\epsilon_\rho/10^{-10}) B_{\chi\to \nu_\tau}\Gamma_\chi \ls 1.5\times 10^{-29}-1.5\times  10^{-28}$ s$^{-1}$. For $\chi\to \nu_\mu \bar{\nu}_\mu$, $m_\chi = 10^8-10^{10}$ GeV and $E_\mu>10^7$ GeV, the range of constraints
  is $(\epsilon_\rho/10^{-10}) B_{\chi\to \nu_\mu}\Gamma_\chi \ls 3\times 10^{-29}-7\times  10^{-28}$ s$^{-1}$.
\end{abstract}
\maketitle

\section{Introduction}

The identification and characterization of dark matter has been a central effort to understand the composition of 27\% of the present day Universe. The weakly-interacting massive particle (WIMP) paradigm, in its simplest form, has weak-scale interacting particles in thermal equilibrium in the early Universe. As the expansion rate of the Universe exceeds their interaction rate, dark matter decouples (``freezes out") at a number density to account for the dark matter fraction of the energy density of the Universe~\cite{Kolb:1990vq,Bertone:2004pz,Feng:2010gw}. Efforts for direct and indirect detection of dark matter yield upper bounds on the WIMP-nucleon cross section as a function of WIMP mass~\cite{Undagoitia:2015gya,Gaskins:2016cha}. Concurrently, LHC searches for WIMP candidates exclude a region of parameter space~\cite{Penning:2017tmb,Rappoccio:2018qxp,Buchmueller:2017qhf}. 

In the most straightforward approach, there is a single species of stable WIMPs. In the WIMP paradigm, if dark matter capture and dark matter annihilation are in equilibrium, for example in the Sun, the annihilation rate is related to the capture rate and  dark matter concentrates at the Sun's core
\cite{Gould:1987ir,Gould:1991hx}. However, equilibrium times for WIMP capture and annihilation in the Earth in the standard WIMP scenario are longer than the age of the solar system \cite{Zentner:2009is,Peter:2009mi}. Indirect constraints on dark matter come from, for example, IceCube searches for dark matter annihilation or in the case of an extended dark sector or small couplings to standard model particles, searches for dark matter decay \cite{IceCube:2016dgk,IceCube:2018tkk,Renzi:2019drf,ANTARES:2020leh}. Constraints come from the absence of signals of annihilation or decay at the center of the Earth, in the core of the Sun and from the dark matter halo in the galaxy.   

As constraints on the simplest WIMP sectors are tightened, theories of a more complex dark sector or long-lived, but unstable, dark matter have been proposed. In some of these models, the relic abundance of dark matter does not follow from the standard picture of thermal equilibrium followed by freeze-out. For example, non-thermal production of super-heavy dark matter (e.g., \cite{Chang:1996vw,Kuzmin:1999zk,Chung:2001cb} and references therein) can evade mass limits that follow from unitarity constraints on cross sections in the thermal freeze-out picture \cite{Griest:1989wd}.  

In setting constraints on dark matter (DM) that accumulates in the Earth \cite{Aartsen:2016fep}, the DM cross section with nuclei, its lifetime and its annihilation cross section enter into the limits. Our focus here is very long-lived, asymmetric superheavy dark matter (SHDM) $\chi$ that comprises a fraction $f_\chi$ of the local DM density $\rho_{\rm DM}$. We consider DM lifetimes of order $\tau\sim 10^{28}$ s, much longer than the age of the Universe. As asymmetric DM, the DM does not (self-)annihilate. With sufficiently large cross sections, the Earth's DM capture rate of SHDM is independent of cross section \cite{Acevedo:2020gro,Mack:2007xj}.
Cosmological and astrophysical considerations permit large dark matter-baryon interaction cross sections (see, e.g., \cite{Boddy:2018kfv,Cyburt:2002uw,Bhoonah:2018gjb,Chivukula:1989cc,Starkman:1990nj,Natarajan:2002cw,Dvorkin:2013cea,Buen-Abad:2021mvc}).

In constraints on dark matter accumulation in the Earth \cite{Aartsen:2016fep}, it is conventionally assumed that dark matter is concentrated at the core of the Earth. With sufficiently large dark matter interactions with baryons in the Earth and absent a significant self-annihilation cross section, the dark matter density at or near the surface of the Earth can be considerably enhanced relative to the local halo density \cite{Neufeld:2018slx}. In ref. \cite{Neufeld:2018slx}, for a dark matter mass equal to twice the proton mass, the dark matter density in the Earth would be nearly uniform, but it becomes more concentrated in the Earth's core for heavier masses. In more complex dark matter scenarios, the drift of heavy dark matter particles to the center of the Earth may be very slow or arrested \cite{Pospelov:2020ktu,Pospelov:2019vuf,Lehnert:2019tuw,Rajendran:2020tmw,Acevedo:2020gro}. Non-standard dark matter density distributions have been proposed~\cite{Anchordoqui:2018ucj} to account for two unusual events with large elevation angles reported by the ANITA collaboration~\cite{Gorham:2016zah,Gorham:2018ydl}. This model would require almost all dark matter particles intercepted by the Earth during its lifetime to be captured~\cite{Anchordoqui:2018qom}. Recent results from ANITA-IV do not show additional events at such large angles~\cite{Gorham:2020zne}. Nevertheless, these theoretical and observational developments emphasize the opportunity for neutrino telescopes to search for dark matter annihilation or decay that originate from a wider density distribution of dark matter throughout the Earth, instead of originating only from the center of the Earth, as is the case in e.g., ref. \cite{Feng:2015hja}.

 In this paper, we consider the case of long-lived superheavy dark matter (SHDM) $\chi$ with mass in the range of $M_\chi=10^7-10^{10}$ GeV. This mass range is chosen to illustrate the capabilities of IceCube 
 to constrain a combination of dark matter density in the Earth and its lifetime in a nearly background-free energy regime. We focus on long-lived, but unstable bosonic dark matter with decays to $\chi\to\nu_\tau \bar\nu_\tau$. We also show results for $\chi\to \nu_\mu \bar\nu_\mu$, and it is straightforward to extrapolate to other decay channels. 
 We take a model independent approach to the origin of the dark matter density profile in the Earth and use two simplified dark matter density profiles: a uniform density and one that scales with the Earth's density of ordinary matter.
We compare IceCube limits on the partial decay width of SHDM decays to neutrinos in the Galactic Center to our limits on the fraction of the Earth's density comprised of dark matter time the partial decay width for DM density distribution profiles that include dark matter far from the Earth's core.

The outline of the paper is as follows. In Sec.~\ref{sec:2}, we briefly discuss our assumptions for the dark matter density distributions in the Earth.
In Sec.~\ref{sec:3}, we describe the evaluation of the number of events using an approximation of IceCube's acceptance for these events.  
Section \ref{sec:4} shows our estimates of the sensitivity of IceCube 
to these SHDM decays. The paper wraps up with discussion and conclusions presented in Sec.~\ref{sec:5}. 

The large dark matter-nucleon cross sections required for appreciable DM densities away from the Earth's core introduce theoretical challenges associated with unitarity constraints \cite{Digman:2019wdm}. While not the main focus of this paper, a discussion of the relation between anomalous DM density distributions in the Earth and the DM cross section with nucleons and nuclei, and the relation of astrophysical and cosmological limits to the cross sections required, is included in the Appendix.

\section{Dark matter in the Earth}
\label{sec:2}

Dark matter accumulation in the Earth depends on the Earth's DM collection efficiency and on DM losses through annihilation and decay. With weak interaction scale DM-nucleon cross sections, DM accumulation is hampered by the Earth's low escape velocity
$v_{\rm esc}=11.2$ km/s, small compared to the average velocity of the Earth relative to the DM background, $v_{\rm rel} \simeq 220~{\rm km/s}$ for a Maxwellian distribution of velocities. This is in contrast to the solar collection efficiency of DM given $v_{\rm esc}=615$~km/s for the Sun.

Proposals of efficient collection of dark matter rely on strong interactions of dark matter.  If the hidden sector
is multi-component and self-interacting, the distribution function is still Maxwellian but the velocity dispersion depends on the mass
spectrum of the hidden sector and could be smaller than
$v_{\rm esc}$~\cite{Foot:2012cs}.
For example, consider a
two-component dark matter model with a large mass hierarchy
$m_a \ll m_b$. The particles can be taken as two heavy sterile
neutrinos charged under a hidden $U(1)$ gauge group. The charges $Q_a$
and $Q_b$ are opposite in sign and satisfy $n_a Q_a + n_b Q_b = 0$,
where $n_a$ and $n_b$ are the number densities of the sterile
neutrinos. In the Galactic halo the dark photon interactions would keep
the sterile neutrinos at a common temperature, and so in the proximity
of the Earth we could have
$f_\oplus (v_b)\propto e^{-v_b^2/v_{0,b}^2}$, with a velocity
dispersion $v_{0,b} \ll v_{\rm rel}$. Thus, in principle, a large mass
hierarchy could allow for both sufficient trapping of particle of type $a$
if $m_p \agt m_a \ll m_b$ and sufficient trapping of particle of type $b$
via self interaction if $v_{o,b} \ll v_{\rm esc}$. It is clear that
even if this toy model can accommodate the data it would require a
large amount of fine-tuning. 

Another approach is to consider DM with strong interactions with nucleons. For $m_\chi\sim 1-10$ GeV with cross sections in the $\sim 10^{-29}-10^{-26}$ cm$^2$ range,  Neufeld, Farrar and McKee \cite{Neufeld:2018slx} have shown that over times much shorter that the age of the Earth $t_\oplus$ = 4.55~Gyr~\cite{Patterson}, the ``hadronically interacting'' dark matter particles equilibrate thermally with the material in the Earth and result in enhancements of the average DM mass density in the Earth relative to its density in the Galactic plane. For $m_\chi\sim $ 1 GeV, the enhancement is $\sim 10^{14}$ times larger than the ambient DM mass density in the Galactic plane,
$\rho_{\rm DM,gal} \simeq 0.5~{\rm GeV}/{\rm cm}^3$ ~\cite{Bienayme:2014kva,Piffl:2014mfa,McKee:2015hwa,Sivertsson:2017rkp} and the DM is distributed throughout the Earth. Dark matter accumulation and slow drift times to the center of the Earth for a range of masses are discussed in ref. \cite{Acevedo:2020gro}. Direct detection experimental constraints on these DM particles do not apply when the DM does not reach the underground detector or is thermalized such that it cannot provide a sufficient interaction recoil energy to be detected. This is discussed in more detail in the Appendix.

Larger DM masses are also considered in a series of papers on milli-charged relics \cite{Pospelov:2020ktu,Pospelov:2019vuf,Lehnert:2019tuw,Rajendran:2020tmw}. Depending on the fraction of DM comprised of virialized milli-charged relics, the milli-charged dark matter density in the Earth relative to the DM mass density in the Galactic plane can be comparably enhanced and may be distributed throughout the Earth. 

In the simplest approach, one can assume that the dark matter distribution is a consequence of large $\chi A$ interactions that yield drift times comparable to the age of the Earth. Accounting for viscous drag, cross sections in the range of $10^{-21}-10^{-14}$ cm$^2 (m_\chi/10^7$ GeV) are required \cite{Acevedo:2020gro}. Unitarity considerations exclude such SHDM as point particles, however, they can be composite particles \cite{Digman:2019wdm}. Drift time limits on the cross sections, the subtleties associated with interpreting constraints on such large cross sections, and how the compare to astrophysical and cosmological constraints are outlined in the Appendix.

Rather than referring to a specific model or scenario of DM interactions in the Earth, we rely on
experiment to resolve the issue of the DM distribution in the Earth. We assume that there is no dark matter annihilation. The evaporation rate of SHDM should be low \cite{Garani:2021feo}.
We discuss here the capability of neutrino telescopes to constrain a scenario of efficient dark matter accumulation by the Earth and slow drifting towards the Earth's core through a constraint on the fraction of the Earth's density comprised of DM.  We focus on long-lived DM which decays to $\nu_\tau+\bar\nu_\tau$ and $\nu_\mu+\bar\nu_\mu$. 

In this paper, we consider dark matter distributed uniformly in the Earth  and a density that mirrors the Earth's mass density. In both cases, our starting point is that the dark matter mass within the Earth accounts for $\epsilon_\rho$ of the Earth's mass. For ease of notation, we define $\rho_\chi\equiv \rho_{\rm DM,Earth}$. With this notation, we consider $\rho_\chi=\epsilon_\rho \rho_\oplus^{\rm avg}$,
and we consider a DM density proportional to the Earth's matter density distribution as parameterized in the Preliminary Earth Reference Model (PREM) \cite{PREM}, $\rho_\chi = \epsilon_\rho \rho_\oplus^{\rm PREM}$.

Efficient capture of DM by the Earth allows $\epsilon_\rho$ to be as large as $\epsilon_\rho\sim  10^{-9}$ \cite{Acevedo:2020gro}. For reference, the uncertainty in the mass of the Earth is  $\epsilon_\rho = 10^{-4}$. The annual net loss of conventional matter from the Earth, primarily hydrogen and helium gas that escapes the atmosphere, is of order $\epsilon_\rho\sim 10^{-17}$ per year.
Kinetic heating from DM capture in the Earth is of order a few times $10^{-3}$ TW, independent of mass for SHDM, and much smaller than the upper bound of 20 TW that comes from estimating the heat flow that is not modeled by radioactivity and processes in the Earth's core, given a total internal heat flow of 44 TW \cite{Mack:2007xj}.

While composite DM is most naturally bosonic, in the case of a multi-particle DM sector with fermion $\chi$, our result can be extended to decays such as $\chi\to \nu H$. In this case, the decay can result in
thermal energy release.  Conservatively, if we assume that the $\chi$ rest mass energy goes into standard model particles upon decay of each DM particle, with the energy given up to heat the Earth, the rate of heating is $ \Gamma_{\rm heat} = m_\chi N_\chi \Gamma_\chi$ where $\tau_\chi=1/\Gamma_\chi$ and $N_\chi$ is the number of DM particles in the Earth. Taking $m_\chi N_\chi=\epsilon_\rho M_\oplus$, the heating rate is 
\begin{equation}
    \Gamma_{\rm heat} = \epsilon_\rho\, \Biggl(\frac{10^{28}\ {\rm s}}{\tau_\chi}\Biggr)\times 54\ {\rm TW}<20\ {\rm TW}\, .
\end{equation}
The constraint below are relevant to $\epsilon_\rho \sim 10^{-10}$ and $\tau_\chi\sim 10^{28}$ s, so the heating rate of the Earth is not relevant to our scenario.

\section{Neutrinos from dark matter decay in the Earth}
\label{sec:3}

\begin{figure}[ht]
%\postscript{AnitonGeometry2.pdf}{0.7}
\includegraphics[width=0.43\textwidth, trim={0 6.5cm 0 5.5cm},clip]{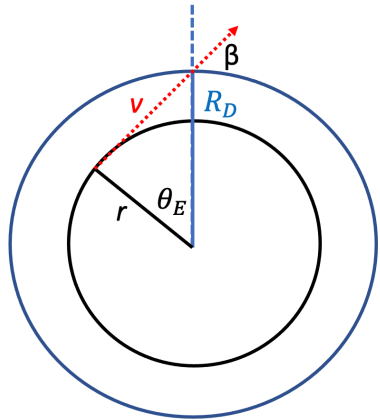}
\caption{Dark matter decay a distance $r$ from the Earth's center, with detector at radial distance $R_D=R_\oplus- d$, where $d$ is the detector distance below the Earth's surface. The distance from the decay point to the detector is denoted $v$.}
\label{fig:geometry}
\end{figure}

We evaluate the number of tau (anti-)neutrino-induced charged leptons produced in  or arriving at a detector a distance $d$ below the surface of the Earth, so a distance $R_D=R_\oplus - d$ from the Earth's center, as shown in Fig.~\ref{fig:geometry}. We assume that $d\ll R_\oplus$. Tau neutrinos come from the decays of DM particles distributed throughout the Earth.

The discussion here focuses on neutrinos. While at lower energies, the neutrino and antineutrino cross sections differ, for the energy range of interest here ($\geq 5\times 10^6$ GeV), the cross sections are essentially equal \cite{Gandhi:1998ri}, so the event rates are also equal. In Sec. IV, all of our results include both neutrino and antineutrino induced events.

We begin with an evaluation of signals of upward-going taus, so DM decays can occur a distance $r\leq R_D$ at angles $\theta_E$ relative to the direction of the nadir of the detector. We label the line-of-sight distance from the detector to the decaying $
\chi$ by $v$, where 
\begin{equation}
\label{eq:v}
   v^2=r^2+R_D^2-2rR_D\cos\theta_E\,. 
\end{equation} 
The total number of $\tau$'s that can be detected depends on the DM number density in the Earth, $n_{\chi}(r)=\rho_{\chi,{\rm Earth}}(r)/m_{\chi}$, the total decay probability over the lifetime $T_0$ of the detector $P_{\chi\to \nu_\tau}(T_0)$, and the probability that the $\tau$ is detected, $P^{\nu_\tau\to \tau}$. The  probability $P^{\nu_\tau\to \tau}$ depends on the location of the DM decay and on the neutrino energy, determined by the mass of the DM particle. 
Suppressing the energy dependence, the number of events is
\begin{equation}
    N_\tau=\int
    d^3 r \ {n_{\chi}(r)}\, P_{{\chi}\to \nu_\tau}(T_0)
    P^{\nu_\tau\to \tau}(r,\theta_E)\frac{\Delta \Omega_{\rm obs}}{4\pi}\ .
    \label{eq:ntau}
\end{equation}
The factor $\Omega_{\rm obs}/(4\pi)$ accounts for the fraction of the isotropic $\chi$ decays that arrive at the detector. For a detector with cross sectional area of radius $r_d$, the fraction  $\chi$ decays that occur a distance
$v$ and have neutrinos that point to the detector is
\begin{equation}
    \frac{\Delta\Omega_{\rm obs}}{4\pi}\simeq \frac{r_d^2}{4 v^2}\, .
\end{equation}
The probability for DM decay to $\nu_\tau$ is
\begin{equation}
    P_{\chi\to \nu_\tau} = B_{\chi\to \nu_\tau}\, \frac{T_0}{\tau_\chi}= B_{\chi\to \nu_\tau}\, \Gamma_\chi T_0\,,
\end{equation}
and  $P_{\chi\to \bar{\nu}_\tau}=P_{\chi\to \nu_\tau} $. Here $B_{\chi\to \nu_\tau}\equiv B(\chi\to \nu_\tau\bar{\nu}_\tau)$.

In what follows, we use an analytic or semi-analytic approximation of  $P^{\nu_\tau\to\tau}$. For $\nu_\tau$ interactions that occur inside the detection region
(labeled with ``start"), for example in IceCube,  $P^{\nu_\tau\to\tau}$ depends on the neutrino charged current cross section, Avogadro's number, the length of the detector $\ell$ and the density of the detector $\rho_{\rm det}$. It also depends on the neutrino attenuation $S(v)$ that depends on the location of the DM decay.  
\begin{equation}
    P^{\nu_\tau\to \tau}\simeq S(v) \sigma_{\rm CC} N_A \ell \rho_{\rm det}\,P_{\rm detect}\, .
    \label{eq:pnutotaustart}
\end{equation}
The neutrino attenuation factor is approximated by
\begin{equation}
    S(v)\simeq \exp(-\sigma_{\rm CC} N_A \rho_{\rm avg,E}\, v)\,, 
\end{equation}
where $v$, defined in eq. (\ref{eq:v}), is a function of $r$ and $\theta_E$. For near surface detectors, almost all of the signal will come from decays in the upper half of the Earth. Consequently, we assign to
$\rho_{\rm avg,E}$ the average density of the Earth along a vertical trajectory from decay point $r$ to $r_{\rm max}$. The quantity $P_{\rm detect}$ is the detection probability. For through-going taus or taus that decay in the detector, 
$P_{\rm detect}$ depends on the tau lifetime, energy and the size of the detector. 

When $\tau$'s are produced outside of the detector, we use approximate formulas for the tau survival probability once it is produced. 
We can reasonably approximate the exit probability as discussed in ref. \cite{Dutta:2005yt}. We assume 
that the distribution of tau energy $E_\tau^i$ from the neutrino charged current interaction
can  be approximated by a $\delta$ function with $E_\tau^i=0.8E_\nu$.  With continuous energy loss characterized by
\begin{equation}
\Biggl\langle \frac{dE_\tau}{dz'}\Biggr\rangle = -b_\tau\rho E_\tau\ ,
\end{equation}
the distance $z'=v-z$ that the tau propagates relates the initial and final tau energy, where the tau is produced at point $z$ along the chord of length $v$. A constant energy loss parameter $b_\tau$ gives a particularly simple form of the probability of the neutrino to produce a tau that arrives at the detector and is detected, for a detector in water or ice, is
\begin{eqnarray}
\nonumber
    P^{\nu_\tau\to \tau}&\simeq &
    \int_{E_\tau^{\rm min}}^{E_\tau^i}\, dE_\tau
    S(z)\,\frac{N_A\sigma_{CC}(E_0)}{b_\tau E_\tau}\\
    &\times & 
    \exp\Biggl[-\frac{m_\tau}{c\tau b_\tau \rho_w}\Biggl(\frac{1}{E_\tau}-\frac{1}{E_\tau^i}
    \Biggr)\Biggr]\, P_{\rm detect}(E_\tau)\, ,
    \label{eq:pexit}
\end{eqnarray}
where $E_\tau^i=0.8 E_0$, $E_\tau^{\rm min}\geq E_\tau^i \exp(-b_\tau\rho_w v)$ for water density $\rho_w$, and the $z= v-\ln(E_\tau^i/E_\tau)/(b_\tau\rho_{\rm avg,E})$ is the
distance the neutrino travels before it interacts to produce a $\tau$ with energy $E_\tau$. 
For the taus that do arrive at a detector in water or ice, they are mainly produced in the last 5--10 km of the neutrino trajectory, which except for near vertical angles, is water. Thus, we use the water density in the exponential written explicitly in eq. (\ref{eq:pexit}).

\begin{figure}[th]
%\postscript{Pexit-anitons-5e8-compare}{0.9}
\includegraphics[width=0.45\textwidth]{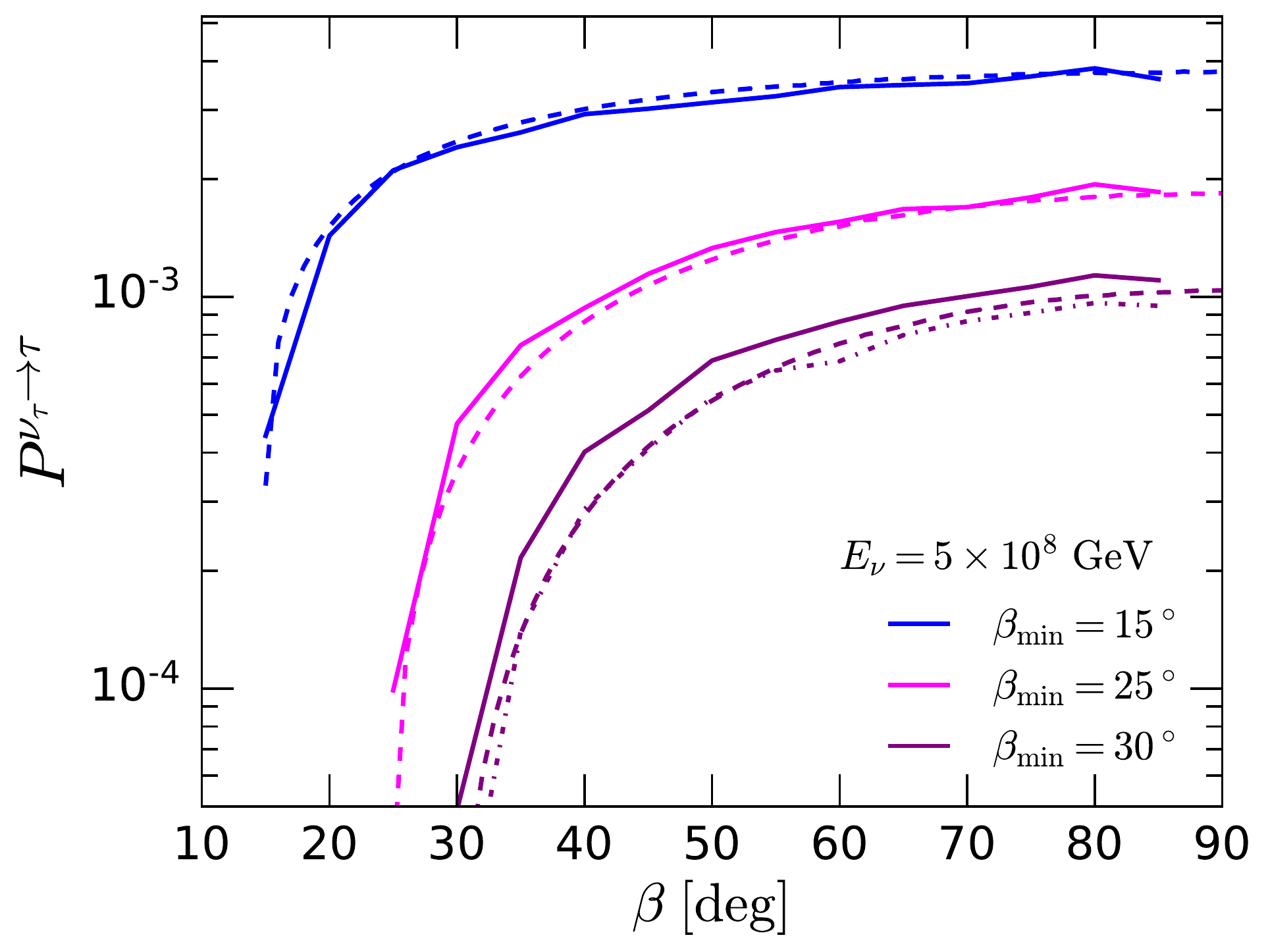}
\caption{$P^{\nu_\tau\to \tau}$ for neutrinos or antineutrinos with energy $5\times 10^8$ GeV, originating
on a shell a distance $r=R_\oplus\cos(\beta_{\rm min})$, as
a function of angle $\beta$ of the trajectory relative to the tangent to the Earth's surface, from Monte Carlo evaluations (solid) and the approximate formula in eq. (\ref{eq:pexit}) (dashed). The dot-dashed line for $\beta_{\rm min}=30^\circ$ shows the Monte Carlo result without regeneration.}
\label{fig:Pexit}
\end{figure}

For our results, we use
\begin{eqnarray}
\nonumber
    b_\tau &=& b_\tau (E_\tau^i)\\
    &=& (1.2+0.16\ln(E_\tau^i/10^{10} {\rm GeV}))\times 10^{-6}\ {\rm cm^2/g}\ .
\end{eqnarray}
for $E_\tau^i>10^8$ GeV and $b=b(10^8\ {\rm GeV})$ for lower energies where, in any case, the decay of the tau dominates over electromagnetic energy loss.

To illustrate how well the approximation of eq. (\ref{eq:pexit}) works, we show a comparison of a Monte Carlo evaluation of
$P^{\nu_\tau\to \tau}$ (solid curves) to the approximate expression (dashed curves) for three shells of DM decays with $R_D=R_E$ with $P_{\rm detect}=1$. In fig. \ref{fig:Pexit},
thin DM density shells are fixed at $r=R_\oplus \cos(\beta_{\rm min})$, where $\beta_{\rm min}=15^\circ$, $25^\circ$ and $30^\circ$ refers to the minimum angle $\beta$ relative to the horizon of an Earth-emerging tau that comes from a shell of DM decays a distance $r$ from the center of the Earth. These angles correspond to DM shells a distance $0.966R_\oplus$, $0.906R_\oplus$ and $0.866R_\oplus$. As the DM decays are deeper in the Earth, tau neutrino regeneration plays a role. For $\beta_{\rm min}=30^\circ$, the dot-dashed curve shows the Monte Carlo result without regeneration processes in which $\nu_\tau\to\tau\to\nu_\tau\to\tau$ in a series of neutrino interactions and tau decays. The analytic approximation does well neglecting regeneration and will yield conservative bounds on the DM matter density fraction in the Earth, $\epsilon_\rho$.

For reference, for muon neutrinos that produce muons in neutrino interactions outside of the detector,
\begin{equation}
    P^{\nu_\mu\to \mu}\simeq 
    \int_{E_\mu^{\rm min}}^{E_\mu^i}\, dE_\mu
    S(z)\,\frac{N_A\sigma_{CC}(E_0)}{b_\mu E_\mu}\\
    \, P_{\rm detect}(E_\mu)\, ,
    \label{eq:pexitmu}
\end{equation}
with the energy loss parameter approximated as
\begin{eqnarray}
\nonumber
    b_\mu&=&b_\mu (E_\mu^i)\\
    &=&(2.2+0.195\ln (E_\mu^i/{\rm GeV}))\times10^{-6}\, {\rm cm^2/g}\,.
\end{eqnarray}

Equation (\ref{eq:ntau}) shows that the unknown parameters, the DM mass fraction of the Earth $\epsilon_\rho$ and the DM lifetime and branching fraction to tau neutrinos come through the combination
\begin{equation*}
    \epsilon_\rho \frac{B_{\chi\to \nu_\tau}}{\tau_\chi} =
    \epsilon_\rho B_{\chi\to \nu_\tau} \Gamma_\chi\,. 
\end{equation*}
We set limits on this combination of unknown parameters based on a lack of tau neutrino plus antineutrino events in this energy range. Corresponding limits are set for muon neutrino plus antineutrino events.

\section{Discovery reach}
\label{sec:4}

\subsection{Events in an IceCube-like detector}

The IceCube detector with $\sim 1$ km$^3$ of instrumented ice can be used to constrain $\epsilon_\rho B_{\chi\to \nu_\tau} \Gamma_\chi$.
We approximate IceCube as an isotropic detector of cross sectional area of $\pi (f_D \times 0.5\ {\rm km})^2$, length $\ell=f_D\times 1\ {\rm km}$. Thus, the fiducial volume is $\sim 0.8 f_D^3$ km$^3$. The average detection efficiency  over the $T_0=10$ years of running time is denoted by $\varepsilon_{\rm det}$.

When signals are through-going tau tracks from taus produced by tau neutrinos and antineutrinos outside of the detector,
the integrand of the probability $P^{\nu_\tau\to\tau}$ in eq. (\ref{eq:pexit}) has
$P_{\rm detect}=\varepsilon_{\rm det} \exp[-\ell/(\gamma c\tau_\tau)]$ accounting for the 
tau lifetime $\tau$ with $\gamma = E_\tau/m_\tau$.
For tracks that convert to showers (decay within IceCube, ``decay'' events), we take $P_{\rm detect} = \varepsilon_{\rm det} (1-\exp 
[-\ell/(\gamma c\tau_\tau)])$ in the integrand of eq. (\ref{eq:pexit}). When neutrinos produce
taus inside the detector, we take $P_{\rm detect}=\varepsilon_{\rm det}$.

For this energy range of $\nu_\tau+\bar{\nu}_\tau$, between $5\times 10^6$
GeV and $5\times 10^{9}$ GeV ($E_{\nu_\tau}=E_{\bar{\nu}_\tau}=E_{0}=m_\chi/2$), we set limits on $\epsilon_\rho B_{\chi\to \nu_\tau} \Gamma_\chi$ assuming $f_D=0.5$ and $\varepsilon_{\rm det}=1$ to illustrate IceCube's capability to constrain the DM density in the Earth. As noted above, we take the DM mass density distributed
uniformly in the Earth according to the Earth's average
density, $\rho_\chi=\epsilon_\rho\rho^{\rm avg}_\oplus=\epsilon_\rho \,(5.5$ g/cm$^3$), 
and a distribution that follows the PREM density, $\rho_\chi = \epsilon_\rho \rho_\oplus^{\rm PREM}$.

\begin{figure}[ht]
%\postscript{IceCube-UniformDensity-Nevt-updatennb}{0.9}
\includegraphics[width=0.45\textwidth]{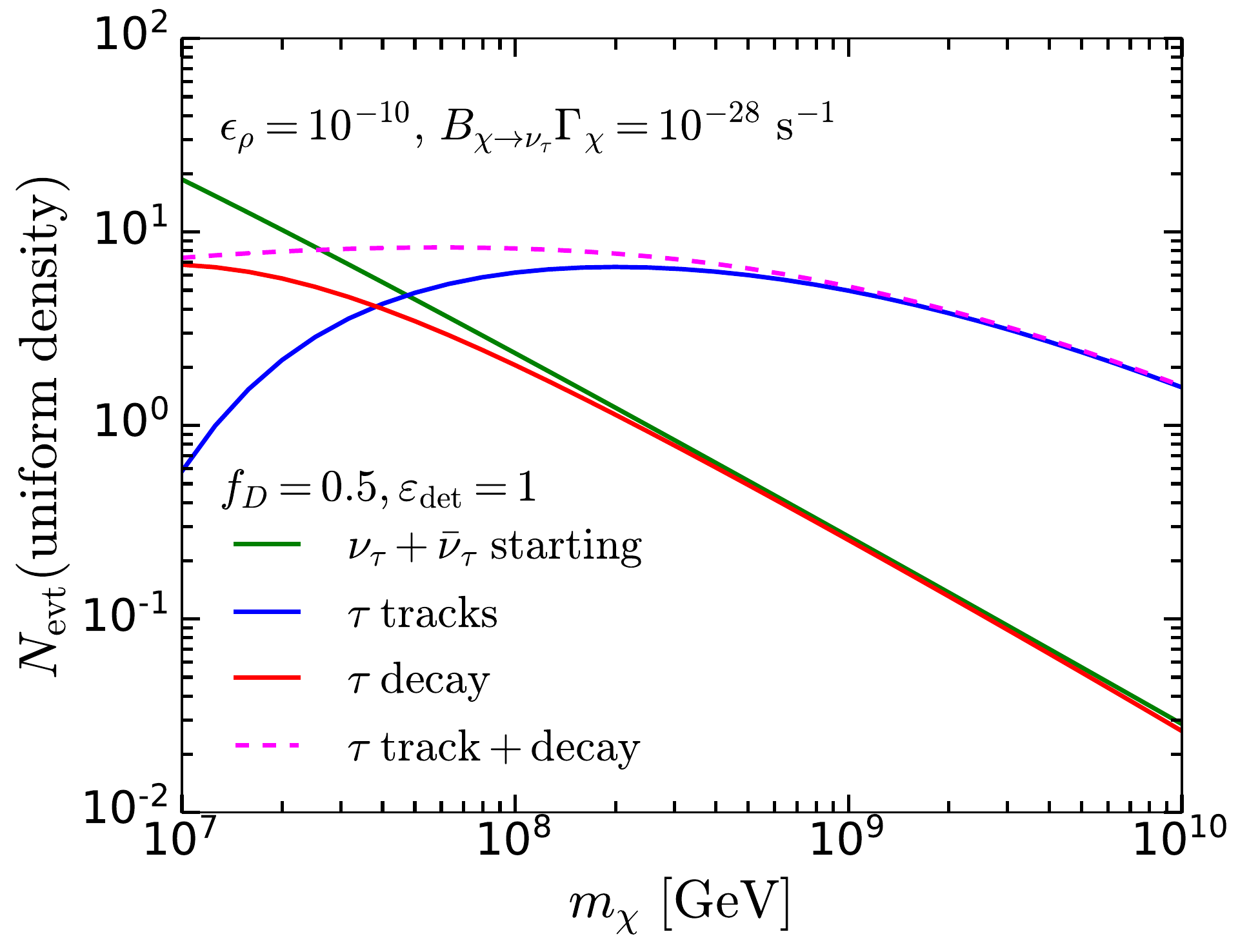}
\caption{The number of events for $\nu_\tau+\bar{\nu}_\tau$ interactions in the detector ($\nu_\tau+\bar{\nu}_\tau$ starting) and $\tau$'s produced outside the detector in ice that pass through the detector ($\tau$ tracks) or decay in the detector ($\tau$ decay). We take $\epsilon_\rho=10^{-10}$ for constant density dark matter $\epsilon_\rho\,(5.5$ g/cm$^3$), $B_{\chi\to \nu_\tau} \Gamma_\chi = 10^{-28}$ s$^{-1}$, detection efficiency $\varepsilon_{\rm det}=1$, fiducial volume fraction $(f_D)^3=(0.5)^3$ and observing time of 10 years.}
\label{fig:nevt}
\end{figure}

As a first demonstration of our results, we show in fig. \ref{fig:nevt} the number of events with these approximations of the detector for 
$\rho_\chi =\epsilon_\rho\, (5.5$ g/cm$^3$), $\epsilon_\rho=10^{-10}$ and $B_{\chi\to \nu_\tau} \Gamma_\chi = 10^{-28}$ s$^{-1}$. The number of events 
has two types of  $m_\chi=2E_0$ dependence. The green curve shows the number of events as a function of $m_\chi$ for events that start with a $\nu_\tau$ or $\bar{\nu}_\tau$ interaction in the detector. The energy dependence associated with the neutrino comes through the neutrino interaction cross section, which grows as $\sim E^{0.3}$ with energy and the neutrino attenuation factor which suppresses the $P^{\nu_\tau\to\tau}$ as energy increases in eq. (\ref{eq:pnutotaustart}). An additional dependence on $v$ in the factor $\Delta\Omega/(4\pi)$ complicates a simple discussion, however, the green line in fig. \ref{fig:nevt} demonstrates the dominant dependence on mass in this case. The number of events scales roughly as $\sim 1/m_\chi$ because we have fixed $n_\chi=\epsilon_\rho\rho_{\rm Earth}/m_\chi$. An increase in the mass of $m_\chi$ has a commensurate decrease in $n_\chi$ that appears in eq. (\ref{eq:ntau}).

With the approximations used here for $P^{\nu_\tau\to \tau}$, neglecting neutrino attenuation, it is straightforward to show that at high energy, the number of events for $\nu_\tau\to \tau$ outside the detector that have tau decays in the detector equals the number of events for the $\nu_\tau$ to produce a tau directly in the detector. The approximate equality holds at high energy as long as the detection efficiencies are equal for $\nu_\tau$ starting and $\tau$ decay events.

The number of events for through-going taus has a different energy behavior than for decaying taus, but the combined number of tau events, shown with the dashed line, is roughly constant at the lower end of the mass scale, then begins to decrease with increasing mass. At low energies, the probability that a tau is produced outside the detector and makes it into the detector scales with the time dilated decay length $\gamma c\tau$, so it increases with $E_0=m_\chi/2$. This directly compensates for the decrease in $n_\chi$ that scales as $1/m_\chi$. For $E_\tau\sim$ a few times $10^8$ GeV, so for $m_\chi$ at twice the scale, the tau range is modified by electromagnetic energy loss 
\cite{Dutta:2000hh} so the range grows more slowly than linearly with energy, thus, the turnover in the number of events.

\begin{figure}[ht]
%\postscript{IceCube-Uniform-all-dr2-nnb}{0.9}
\includegraphics[width=0.45\textwidth]{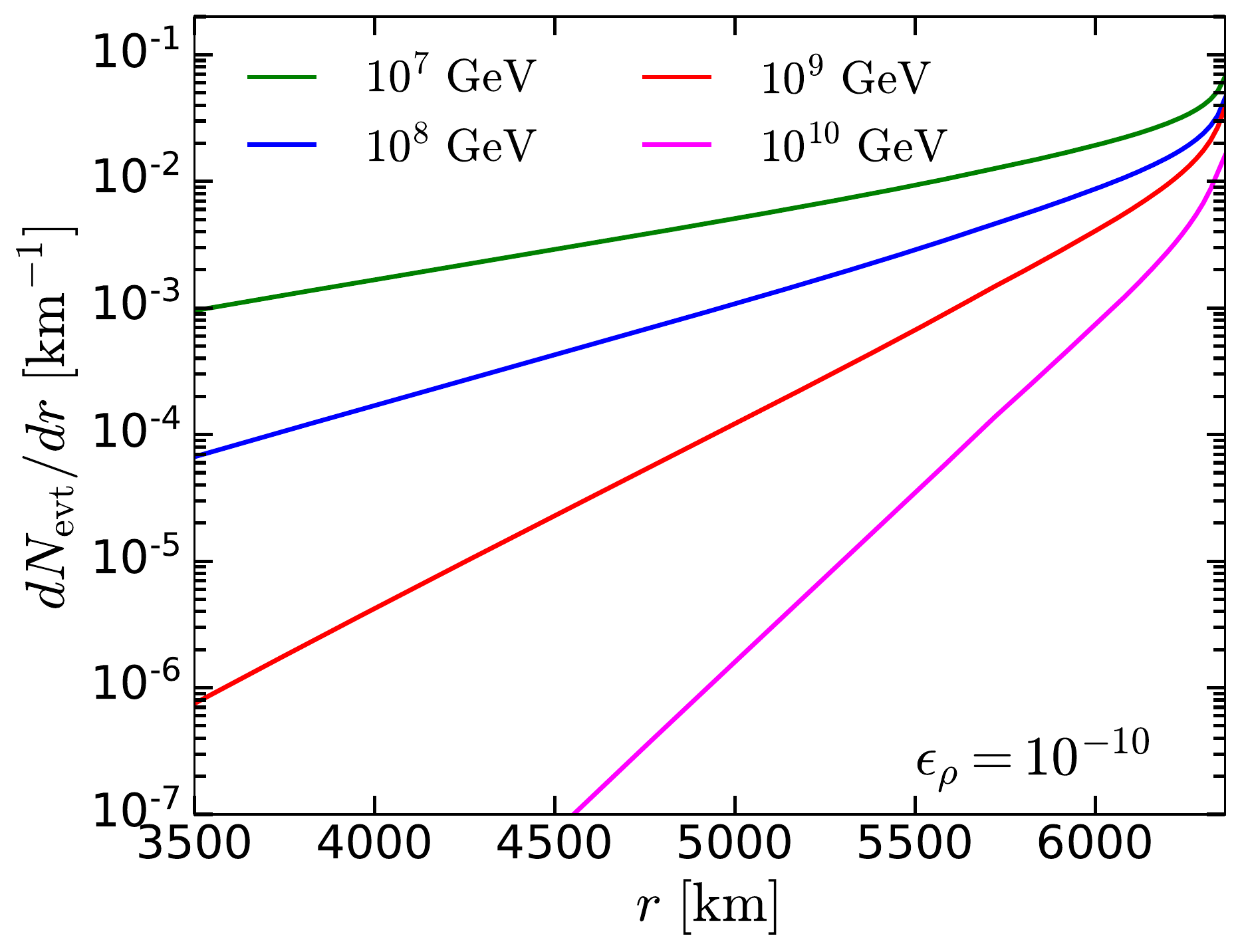}
\caption{The number of events for $\nu_\tau+\bar{\nu}_\tau$ interactions in the detector and tau tracks and decays, all combined, from $\chi\to \nu_\tau+\bar{\nu}_\tau$ as a function of radius for $\epsilon_\rho=10^{-8}$ and for constant density dark matter $\epsilon_\rho\,(5.5$ g/cm$^3$), $B_{\chi\to \nu_\tau} \Gamma_\chi = 10^{-28}$ s$^{-1}$, detection efficiency $\varepsilon_{\rm det}=1$, fiducial volume fraction $(f_D)^3=(0.5)^3$ and observing time of 10 years.}
\label{fig:dndr}
\end{figure}

\begin{figure}[ht]
%\postscript{IceCube-Uniform-all-dangle2-nnb}{0.9}
\includegraphics[width=0.45\textwidth]{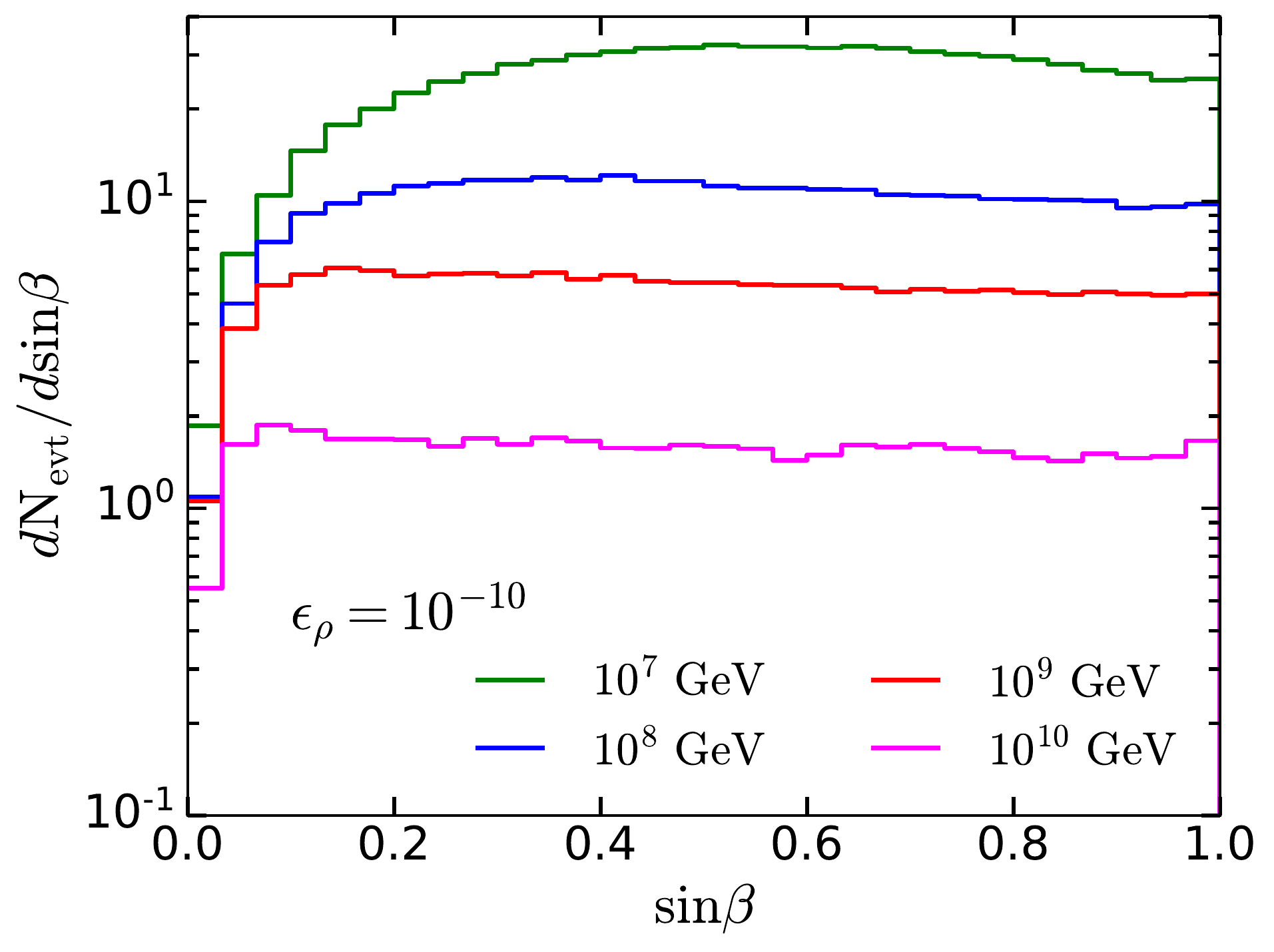}
\caption{The number of events for $\nu_\tau+\bar{\nu}_\tau$ interactions in the detector and tau tracks and decays, all combined, from $\chi\to \nu_\tau+\bar{\nu}_\tau$ as a function of $\sin\beta$ for $\epsilon_\rho=10^{-10}$ and for constant density dark matter $\epsilon_\rho\,(5.5$ g/cm$^3$), $B_{\chi\to \nu_\tau} \Gamma_\chi = 10^{-28}$ s$^{-1}$, detection efficiency $\varepsilon_{\rm det}=1$, fiducial volume fraction $(f_D)^3=(0.5)^3$ and observing time of 10 years.}
\label{fig:dndsinbeta}
\end{figure}

Figures \ref{fig:dndr} and \ref{fig:dndsinbeta} show the distribution of events coming from dark matter $\chi \to \nu_\tau \bar{\nu}_\tau$ decays as a function of radial distance and $\sin\beta$ where $\beta$ the angle of the trajectory relative to the tangent to the Earth's surface. We have only considered upward events. Because of neutrino attenuation, the total number of events (starting, tracks and decays) decreases with energy. Higher energies are dominated by $\chi$ decays closer to the detector. Figure \ref{fig:dndsinbeta} shows that except for nearly horizontal incident particles, the angular distribution is nearly isotropic for the upward event rate. 

\begin{figure}[ht]
%\postscript{IceCube-combined2-2-nnb}{0.9}
\includegraphics[width=0.45\textwidth]{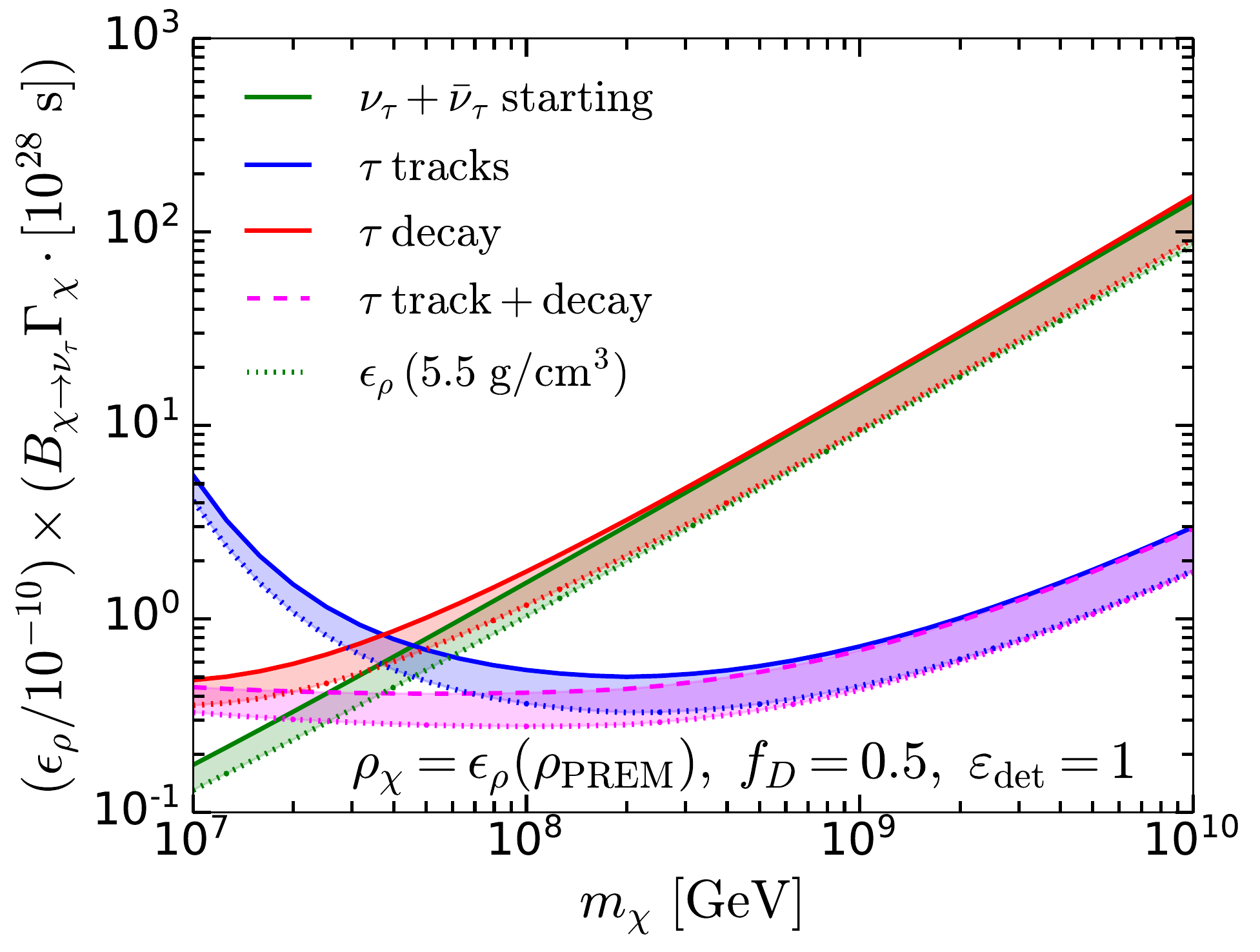}
\caption{Constraints on the $(\epsilon_\rho/10^{-10})\times
B_{\chi\to\nu_\tau}\Gamma_\chi [10^{28}\ {\rm s}]$ for constant density dark matter $\epsilon_\rho\times 5.5$ g/cm$^3$ (dotted curves) and for the dark matter density that is proportional to the PREM density model of the Earth (solid and dashed curves) for an isotropic detector in ice with a fiducial volume of $f_D^3\times 0.8\ {\rm  km}^3$ for $f_D=0.5$ and detection efficiency $\varepsilon_{\rm det}=1$. The neutrino and antineutrino energy that produces these events is assumed to be $E_0=m_\chi/2$ in $\chi\to \nu_\tau\bar{\nu}_\tau$. Allowed parameters are below the curves in the figures. An observing time of 10 years is assumed. }
\label{fig:IceCube}
\end{figure}

An exclusion region for $\epsilon _\rho B_{\chi\to \nu_\tau}\Gamma_\chi$ that is achievable by an IceCube-like detector is shown in fig. \ref{fig:IceCube}. It is obtained by setting $N_{\rm evt}=2.44$, the number of events associated with a 90\% CL limit assuming no background. The region below the solid curves for each of the detection channels is allowed, for example, for $B_{\chi\to \nu_\tau} \Gamma_\chi = 10^{-28}$ s$^{-1}$, for $m_\chi=10^8$ GeV, 
$\epsilon_\rho\ls 10^{-10}$ if the distribution of dark matter in the Earth mirrors the Earth matter density, scaled by $\epsilon_\rho$. For the low mass region, the tau events will appear as starting events or decays in the detector. At higher masses, the high energy tau track will have a similar energy loss profile in the detector as lower energy muons. We have $b_\tau/b_\mu\simeq 0.08$ for charged lepton energy $10^8$ GeV, so a $10^8$ GeV tau will appear to be a $\sim 8\times 10^6$ GeV muon of the basis of $\langle dE_\tau/dX\rangle$. 
Better modeling of muon stochastic energy losses \cite{Abbasi:2021czc} may ultimately help distinguish high energy taus from high energy muons.

\begin{figure}[ht]
%\postscript{IceCube-combined-layers2-nnb}{0.9}
\includegraphics[width=0.45\textwidth]{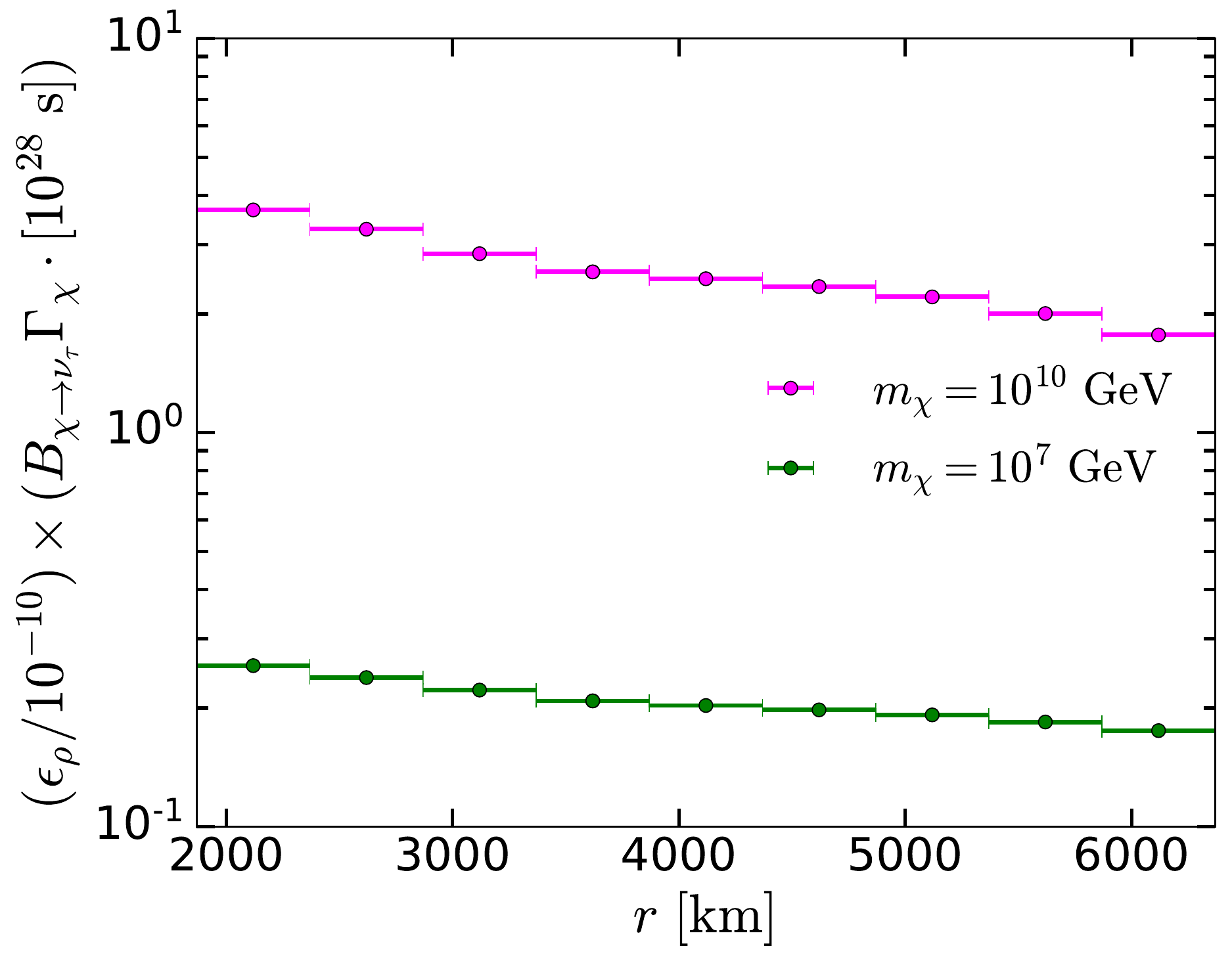}
\caption{For two DM masses, constraints on the $\epsilon_\rho\times
B_{\chi\to \nu_\tau}\Gamma_\chi$
as in fig. \ref{fig:IceCube} for constant density dark matter $\epsilon_\rho\times 5.5$ g/cm$^3$ for DM in radial shells, for an isotropic detector in ice with a volume of $f_D^3\times 0.8\ {\rm  km}^3$ for $f_D=0.5$ and detection efficiency $\varepsilon_{\rm det}=1$. The neutrino and antineutrino energy that produces these events is assumed to be $E_0=m_\chi/2$ in $\chi\to \nu_\tau\bar{\nu}_\tau$. Allowed parameters are below the curves in the figures. An observing time of 10 years is assumed. }
\label{fig:IceCube-layers}
\end{figure}

The density distribution of DM in the Earth is unknown. Figure \ref{fig:IceCube-layers} shows the potential constraints on the contributions to a tau neutrino signal from a 500 km shell of DM. For each 500 km interval in $r$, a predicted number of events of 2.44 sets the limits shown in the figure.

\begin{figure}[ht]
%\postscript{IceCube-combined2-muons2-1e7-nnb}{0.9}
\includegraphics[width=0.45\textwidth]{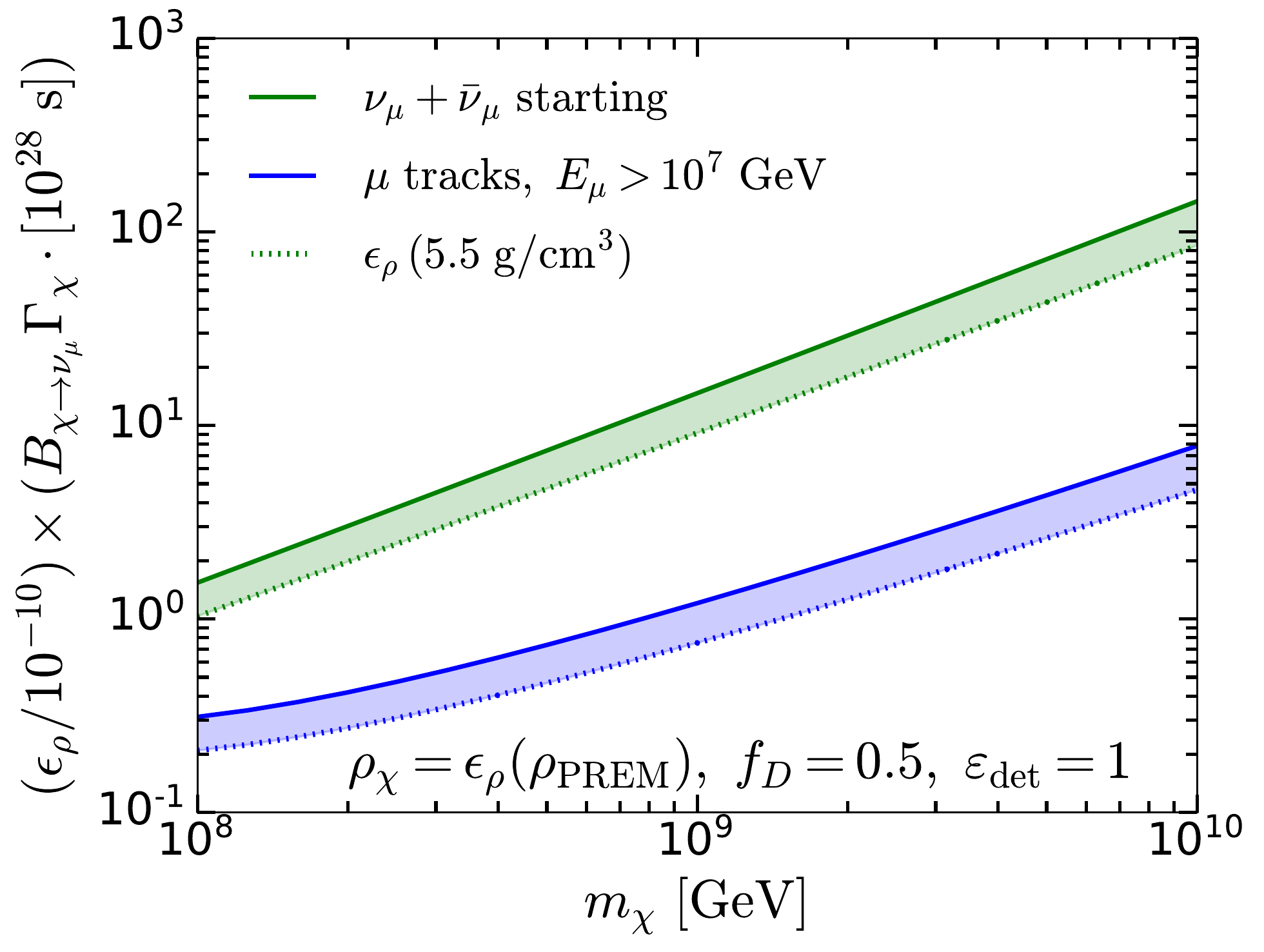}
\caption{Constraints on the $(\epsilon_\rho/10^{-10})\times
B_{\chi\to\nu_\mu}\Gamma_\chi [10^{28}\ {\rm s}]$ for constant density dark matter $\epsilon_\rho\times 5.5$ g/cm$^3$ (dotted curves) and for the dark matter density that is proportional to the PREM density model of the Earth (solid and dashed curves) for an isotropic detector in ice with a fiducial volume of $f_D^3\times 0.8\ {\rm  km}^3$ for $f_D=0.5$ and detection efficiency $\varepsilon_{\rm det}=1$. The neutrino and antineutrino energy that produces these events is assumed to be $E_0=m_\chi/2$ in $\chi\to \nu_\mu\bar{\nu}_\mu$. Allowed parameters are below the curves in the figures. An observing time of 10 years is assumed, and the muon energy at the detector is required to be above $10^7$ GeV. }
\label{fig:IceCube-muons}
\end{figure}

\begin{figure}[ht]
%\postscript{IceCube-combined-layers-muons2-1e7-nnb}{0.9}
\includegraphics[width=0.45\textwidth]{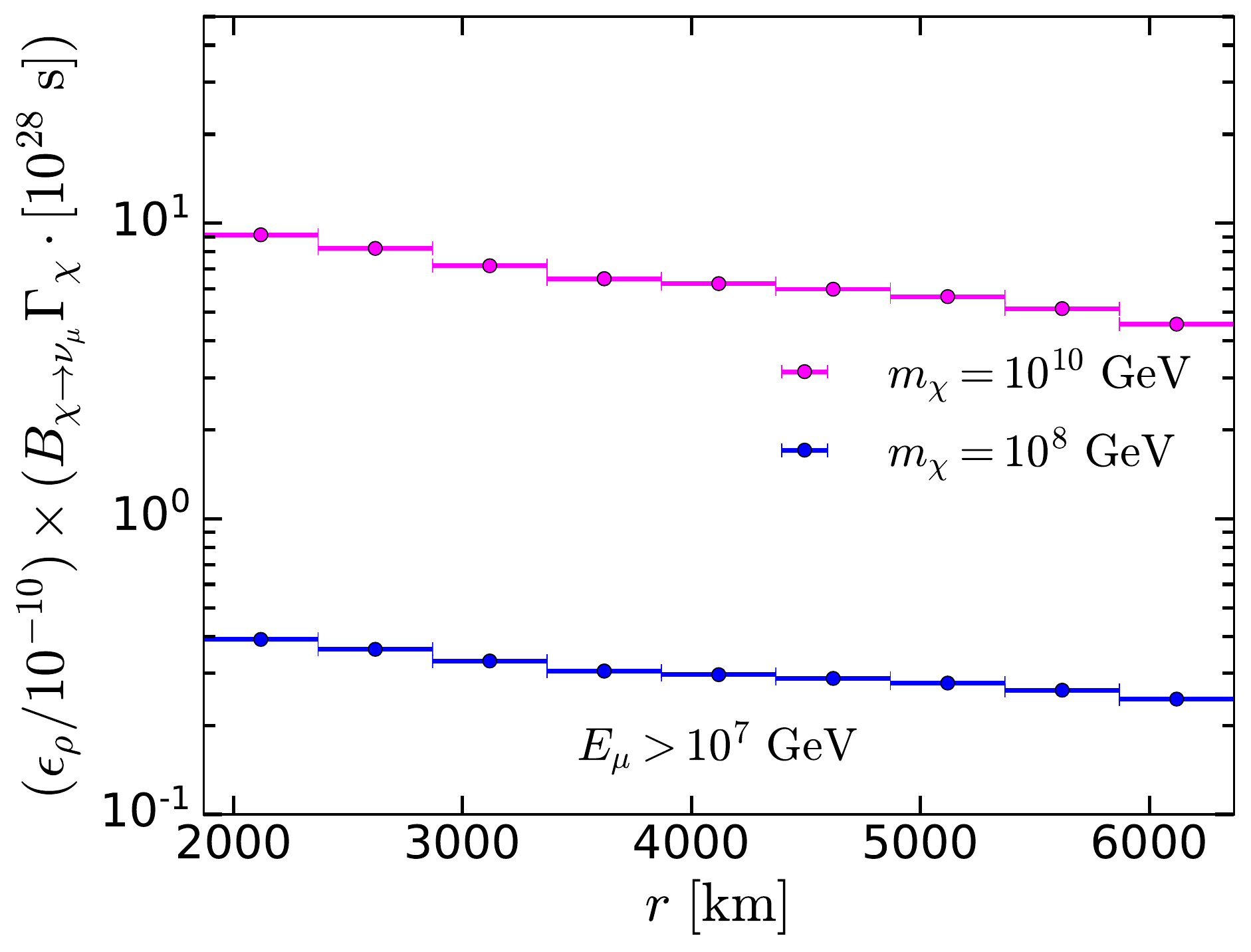}
\caption{For two DM masses, constraints on the $\epsilon_\rho\times
B_{\chi\to \nu_\mu}\Gamma_\chi$
as in fig. \ref{fig:IceCube-muons} for constant density dark matter $\epsilon_\rho\times 5.5$ g/cm$^3$ for DM in radial shells, for an isotropic detector in ice with a volume of $f_D^3\times 0.8\ {\rm  km}^3$ for $f_D=0.5$ and detection efficiency $\varepsilon_{\rm det}=1$. The neutrino and antineutrino energy that produces these events is assumed to be $E_0=m_\chi/2$ for $\chi\to \nu_\mu \bar{\nu}_\mu$. Allowed parameters are below the curves in the figures. An observing time of 10 years is assumed, and $E_\mu>10^7$ GeV at the detector. }
\label{fig:IceCube-layers-muons}
\end{figure}

We show the corresponding results for $\chi\to \nu_\mu \bar{\nu}_\mu$ in figs. \ref{fig:IceCube-muons} and \ref{fig:IceCube-layers-muons}. We set
$E_\mu^{\rm min}=10^7$ GeV to avoid an energy region with background events. The muon track sensitivity is better than the $\nu_\mu$ starting event sensitivity because the muon range is larger that $f_D\ell=0.5$ km for the whole range of $m_\chi$ shown. For larger $m_\chi$, a higher energy muon neutrino emerges, which in turn produces a higher energy muon with a longer range. The muon range scales as $\sim 1/b_\mu\cdot \ln (E_\mu^i/E_\mu^{\rm min})$. The sensitivity curves are slightly lower for $\chi\to \nu_\mu \bar{\nu}_\mu$ than 
for $\chi\to \nu_\tau \bar{\nu}_\tau$ for $m_\chi=10^8$ GeV because the muon range is larger than the tau range at these energies. For high $m_\chi$, the tau range is longer. The relative sensitivities for layers of constant density dark matter in the Earth for $\chi\to \nu_\mu \bar{\nu}_\mu$ compared to $\chi\to \nu_\tau \bar{\nu}_\tau$ also shows this effect, as seen in the comparison of the results shown for $m_\chi=10^{10}$ GeV in figs. \ref{fig:IceCube-layers} and \ref{fig:IceCube-layers-muons}.

\subsection{Constraints from the Galactic center observations}

IceCube observations of the Galactic center (GC)   provide an alternative probe of SHDM decay~\cite{Aartsen:2018mxl}. If the lifetime of the SHDM particle is longer than the age of the Universe, $\tau_\chi > t_{\rm U}$, the differential $\nu + \bar \nu$ flux per flavor from SHDM decay in a cone of half-angle $\psi$ around the GC, covering a field of view $\Delta \Omega = 2 \pi (1 - \cos \psi)$, is given by
\begin{equation}
\frac{d\Phi}{dE_\nu} = \frac{\Delta \Omega}{4\pi} \ {\cal J}_{\Delta \Omega} \ \frac{R_{\rm sc} \ \rho_{\rm DM,gal}  \ \Gamma_{\chi \to \nu \bar{\nu}}}{m_\chi} \ \frac{1}{3} \ \frac{dN}{dE_\nu}
\end{equation}
where $dN/dE_\nu = 2\delta(m_\chi/2 - E_\nu)$ is the $\nu + \bar \nu$ spectrum produced per decay, $R_{\rm sc} = 8.5~{\rm kpc}$ is the solar radius circle, $\Gamma_{\chi \to \nu \bar\nu}$ is the  partial decay width to three flavors, $\rho_{\rm DM,gal}$
is the normalizing DM density introduced in Sec.~\ref{sec:2} (which is equal to the commonly quoted DM density at $R_{\rm sc}$) and ${\cal J}_{\Delta \Omega}$ is the average in the field of view (around the GC) of the line-of-sight integration of the DM density, which is found to be
\begin{equation}
{\cal J}_{\Delta \Omega} = \frac{2 \pi}{\Delta \Omega} \ \frac{1}{R_{\rm sc} \ \rho_{\rm DM,gal}}  \int_{\cos \psi}^1 \int_0^{l_{\rm max}} \rho(r) \ dl \ d (\cos \psi') \, ,
\label{Jint}
\end{equation}
where  $l_{\rm max} = \sqrt{R_{\rm halo}^2 - R_{\rm sc}^2 \sin^2 \psi} + R_{\rm sc} \cos \psi$ and $\rho_{\rm DM} (r)$ is the DM density as a function of the distance from the GC,
with $r = \sqrt{R_{\rm sc}^2 - 2 l R_{\rm sc} \cos \psi' +l^2}$~\cite{PalomaresRuiz:2007ry}. For $R_{\rm halo} \agt 10~{\rm kpc}$, eq. (\ref{Jint}) barely depends on the size of the halo. The IceCube Collaboration adopted as benchmark a dark matter distribution which follows the Burkert halo profile~\cite{Burkert:1995yz} with best-fit parameters from~\cite{Nesti:2013uwa}. Other halo profiles (e.g. Navarro- Frenk-White~\cite{Navarro:1995iw}) were considered as systematic uncertainties. An in-depth analysis of the uncertainty in the DM distribution and the implications for the potential sensitivity to DM annihilation or decay in the GC appears in ref.  ~\cite{Guepin:2021ljb}. 

The null result from IceCube searches can be used to set an upper bound on $d\Phi/dE_\nu$, which can be translated into an upper bound on the partial decay width  as a function of $m_\chi$.  The  limit from IceCube null results is approximately $B_{\chi\to \nu \bar\nu}\Gamma_\chi \alt 1-2\times 10^{-28}~{\rm  s}^{-1}$~\cite{Aartsen:2018mxl} for $m_\chi\simeq 10^7$ GeV, with the exception at the energy of the $\bar\nu_e$ Glashow resonance, where the limit is an order of magnitude better.  
From the non-observation of upward going neutrinos originated in the decay of SHDM trapped inside the Earth shown in fig. 6, for $m_\chi=10^7$ GeV the limit is $(\epsilon_\rho/10^{-10})\times (B_{\chi\to \nu_\tau}\Gamma_\chi )\alt 2\times 10^{-29}~{\rm  s}^{-1}$. For a nominal value of $\epsilon_\rho=10^{-10}$, this gives $B_{\chi\to \nu_\tau}\Gamma_\chi \alt 2\times 10^{-29}~{\rm  s}^{-1}$,
comparable to or better than IceCube's GC all flavor limits.  
Our calculation of starting events when $\nu_\tau$ interacts in the detector applies equally well to starting events for $\chi\to \nu_e \bar\nu_e$ and $\chi\to \nu_\mu \bar\nu_\mu$. Limits from muon tracks from $\chi\to \nu_\mu \bar\nu_\mu$ are more stringent than from starting events. For $m_\chi=10^8$ GeV, $(\epsilon_\rho/10^{-10})\times (B_{\chi\to \nu_\mu}\Gamma_\chi )\alt 3\times 10^{-29}~{\rm  s}^{-1}$. The limit weakens with $m_\chi$ according to $\sim (m_\chi/{10^8\ \rm GeV})^{0.7}$
Should SHDM signals from the GC be observed, $\epsilon_\rho$ can be extracted.
\section{Discussion and Summary}
\label{sec:5}

While not achieved with a specific model, recent interest in DM models that predict that heavy DM will not immediately sink to the center of the Earth have motivated our investigation of  signals of dark matter decay from a non-standard DM density profile in the Earth. Achieving such a distribution of DM throughout the Earth requires large DM cross sections with nuclei, a topic that has been re-opened \cite{Digman:2019wdm} as discussed in the Appendix. While challenging to address astrophysical and cosmological constraints on cross sections together with requirements for DM drift time to the center of the Earth, model dependence leaves open the possibility for unusual DM density profiles. Anomalous events such as those observed by ANITA  \cite{Gorham:2016zah,Gorham:2018ydl} prompt a more expansive consideration of where DM may reside within the Earth. 
We have investigated the potential for an underground detector like IceCube to constrain the fraction of the Earth's matter density comprised of DM, distributed throughout the Earth, in a scenario where DM has a long lifetime and has a two-body decay to a neutrino and antineutrino, with our main focus on tau neutrino signals.  

Our assumptions that
the distribution of DM mass is either a fixed fraction of the Earth's density as a function of radius or a uniform density yield similar results. The limits on  $(\epsilon_\rho/10^{-10}) B_{\chi\to \nu_\tau}\Gamma_\chi$ are within a factor of $\sim 1.6$ of each other. The characteristic constraints lie in the range of $(\epsilon_\rho/10^{-10}) B_{\chi\to \nu_\tau}\Gamma_\chi \ls 3\times 10^{-29}-3\times  10^{-28}$~s$^{-1}$ for 
$m_\chi\sim 10^7-10^{10}$ GeV. 
For a DM mass range of $10^8-10^{10}$ GeV, we find that the constraints range between $(\epsilon_\rho/10^{-10}) B_{\chi\to \nu_\mu}\Gamma_\chi \ls 6\times 10^{-29}-1.4\times  10^{-27}$~s$^{-1}$.
This paper proposes a starting point for measurements of upward-going events, complementary to Galactic center observations, to constrain $\epsilon_\rho\times B_{\chi\to \nu}\Gamma_\chi$.

\acknowledgements
We thank J. Beacom and E. Mayotte for helpful discussions.  This work is supported in part by U.S. Department of Energy Grants DE-SC-0010113 (MHR), DE-SC-0009913 (IS), U.S. National Science Foundation NSF Grant PHY-2112527 (LAA),
  the National Aeronautics and Space Administration NASA Grants
  80NSSC18K0464 (LAA, TP), 80NSSC18K0246 (JE, AVO), 80NSSC19K0626 (JFK), 17-APRA17-0066 (TMV), 
  the Neil Gehrels Prize Postdoctoral Fellowship (CG), 
 and the  Fonds  de  la  Recherche  Scientifique-FNRS,  Belgium,
grant No.~4.4503.19 (AB).

\appendix*

\section{Dark matter-nucleon cross section connections}

We have taken a model independent approach in our analysis to constrain $\epsilon_\rho\times B_{\chi\to\nu_\ell}\Gamma_\chi$ with DM decays distributed throughout the Earth. The quantity $\epsilon_\rho$ depends on the capture rate of DM by the Earth. While there are other more complicated scenarios of DM accumulation that can depend on a complex DM sector as well as DM interaction with nucleons, in this appendix, we discuss the status of constraints on large DM cross sections that arise when the dominant accumulation of DM is  via DM scattering with nucleons and nuclei.

As noted in Section II, large cross sections are required to slow the drift of DM to the center of the Earth, thus allowing a distribution of DM outside of the core of the Earth. For large cross sections, the capture rate by the Earth is independent of cross section. 
For a local dark matter density $\rho_{\rm DM}=0.3$ GeV/cm$^3$ and fraction of local dark matter density that is super heavy,$f_\chi\equiv \rho_\chi/\rho_{\rm DM}$, the capture rate of DM by the Earth is
\cite{Acevedo:2020gro,Mack:2007xj}
\begin{equation}
\label{eq:captureA}
    C_\chi^\oplus \simeq 2.45\times 10^{25}\ {\rm s^{-1}}
    \frac{{\rm GeV}}{m_\chi}\,f_\chi\,.
\end{equation}
The SHDM accumulation in the Earth 
over the age of the Earth
$t_\oplus\simeq 4.5\times 10^9$ yr, with the very long DM lifetimes considered here, yields
\begin{eqnarray}
\nonumber
    M_\chi^\oplus &=& f_\chi \, C_\chi^\oplus \, t_\oplus \, m_\chi 
    \simeq f_\chi\bigl( 3.5\times 10^{42}\ {\rm GeV}\bigr)
    \\ &\simeq & f_\chi \bigl(6\times 10^{15}\ {\rm kg}\bigr)\,.
\end{eqnarray}
This translates to a cross section and lifetime independent result
\begin{equation}
    \epsilon_\rho\simeq f_\chi\times 10^{-9}
\end{equation}
given the Earth's mass, $M_\oplus=5.97\times 10^{24}$ kg, as long as the $\chi$ cross section with nucleons and nuclei is sufficiently large and the lifetime is long compared to $t_\oplus$. Our results which can be scaled to account for any $f_\chi$ are normalized by $\epsilon_\rho=10^{-10}$ where $f_\chi=0.1$.

\begin{table}[]
    \centering
    \begin{tabular}{|c|c|}
    \hline
    Notation & Application for SHDM \\
    \hline \hline
     $\sigma_{\chi N}$    & Related to $\sigma_{\chi A}$ by $\sigma_{\chi A}\simeq A^4\sigma_{\chi N}$ \\ \hline
     $\sigma_\chi$    & Free nucleons $\sigma_\chi=\sigma_{\chi p}=\sigma_{\chi n}$ \\ \hline
     $\sigma_c$ & $A$ independent $\sigma_c=\sigma_{\chi A} = \sigma_{\chi p}$\\
     \hline
    \end{tabular}
    \caption{Notation to distinguish different treatments of SHDM cross sections with nuclei and free nucleons. Here, the reduced mass of the $\chi A$ system is $\sim A m_p$ for proton mass $m_p$.}
    \label{tab:cross_sections}
\end{table}

The cross sections required to achieve a cross section independent capture rate depend on how $\chi$ interactions with nuclei depend on mass number $A$. The isotope dependent, spin independent cross sections for SHDM scattering with nucleus $A$ are usually assumed to scale to with a power of $A^4$ relative to the $\chi$-nucleon cross section. For the definition of $\sigma_{\chi N}$ according to $\sigma_{\chi A}=A^4\sigma_{\chi N}$, eq. (\ref{eq:captureA}) is valid for 
\cite{Acevedo:2020gro,Mack:2007xj}
\begin{equation}
 \sigma_{\chi N}
    \gs 6\times 10^{-32}\ {\rm cm^2} \Biggl( \frac{m_\chi}{10^7\ {\rm GeV}}\Biggr)\,.
\end{equation}
An important component of the capture rate is DM scattering with iron 
$(A=56)$, so the $\sigma_{\chi N}$ is increased by a factor of $\sim 10^7$ to get $\sigma_{\chi A}$.

It has been emphasized, however, that for large 
$\sigma_{\chi N}$ and $\sigma_{\chi A}$, the DM particle cannot be treated as a point particle \cite{Digman:2019wdm}. In particular, the assumed model independent scaling relation $\sigma_{\chi A}\propto A^4\sigma_{\chi N}$ for contact interactions breaks down.
For $10^{-31}\ {\rm cm^2}\gs \sigma_{\chi N}\gs 10^{-25}$ cm$^2$, the scaling of $\sigma_{\chi N}$ constraints 
on lower cross sections to higher cross sections is uncertain, and for cross sections larger than $10^{-25}$
cm$^2$, the DM cannot be a point particle \cite{Digman:2019wdm}. 

Composite dark matter with a physical scale larger than the nuclear size is one possibility to overcome the limitations on $\sigma_{\chi N}$. With large intrinsic radii, composite dark matter cross sections with nuclei could be independent of $A$. Following ref. \cite{Digman:2019wdm}, we consider an $A$ independent SHDM cross section. We denote this composite dark matter cross section with nuclei and nucleons by $\sigma_c=\sigma_{\chi N}=\sigma_{\chi A}$. 
The condition on an isotope independent cross section such that eq. (\ref{eq:captureA}) applies is \cite{Acevedo:2020gro}
\begin{equation}
{\sigma_c}
     \gs 4\times 10^{-25}\ {\rm cm^2}\Biggl( \frac{m_\chi}{10^7\ {\rm GeV}}\Biggr) \,.    
\end{equation}

As emphasized in ref. \cite{Digman:2019wdm}, the scaling with $A$ of a composite dark matter cross section with nuclei can be model dependent. Constraining a specific composite dark matter model is beyond the scope of this paper. Instead, we use the $A^4$ dependent result for $\sigma_{\chi N}$ and the $A$ independent result $\sigma_c$ to bracket the range of DM cross sections with nucleons and nuclei in the following discussion. We also make the distinction of scattering of SHDM with the protons in hydrogen, or with neutrons in the early Universe. The cross section with free nucleons will be denoted by $\sigma_\chi$. 
Table \ref{tab:cross_sections} lists our notation to distinguish SHDM interactions with nucleons/nuclei.

To achieve significant SHDM densities outside of the core of the Earth, cross sections even larger than the lower limits in eqs. (A.4) and (A.5) are required.
The cross section of DM with nuclei determines the drift velocity, so $\sigma_{\chi N}$ and $\sigma_c$ can be constrained.
For large cross sections, the DM drift towards the center of the Earth is strongly affected by a viscous drag force. Setting  the drift time 
$t_{\rm drift}\gs 10^9$ yr gives \cite{Acevedo:2020gro}
\begin{eqnarray}
\label{eq:sigdrift}
\sigma_{\chi N} &\gs& 10^{-21} \ {\rm cm^2}\ \Biggl(\frac{m_\chi}{10^7\ {\rm GeV}}\Biggr)\,,
\end{eqnarray}
for the isotope dependent cross section with nucleons scaled to nuclei, assuming drift times in the Earth's core dominate. The isotope independent limit comes from multiplying by $A^4$ so  
\begin{eqnarray}
\label{eq:sigcdrift}
\sigma_c &\gs &  10^{-14}\ {\rm cm^2}\ \Biggl(\frac{m_\chi}{10^7\ {\rm GeV}}\Biggr)\,.
\end{eqnarray}
Thus, equations (\ref{eq:sigdrift}) for $\sigma_{\chi N}$ and (\ref{eq:sigcdrift}) for $\sigma_{c}$ should bracket the minimum SHDM cross sections with protons, $\sigma_\chi$.

At first sight, large $\sigma_{\chi N}$ and $\sigma_{c}$ appear to be excluded by cosmological and astrophysical constraints and by terrestrial experiments. However, many of the constraints come assuming long $\chi A$ interaction lengths.
For example, cross section limited by eq. (\ref{eq:sigcdrift}) would imply that SHDM interaction lengths are very short,
independent of material composition since the target number density scales as $1/A$ and $m_A=A m_N$.  
Limits on SHDM from underground experiments will not apply, as
SHDM thermalization in the atmosphere means they cannot not provide the recoils needed for detection in underground experiments. For smaller (but still large) cross sections $\sigma_c$ where DM energy loss is not complete, constraints from shallow underground detectors have recently been reported
\cite{Cappiello:2020lbk}, but they are limited by SHDM energy loss in the atmosphere and the 6 m.w.e. of cement above the detector. Attenuation of SHDM in the atmosphere does not permit them to make constraints for $\sigma_c\gs 10^{-20}\ {\rm cm^2}(m_\chi/10^7\ {\rm GeV})$ \cite{Cappiello:2020lbk}. 

We focus the remaining discussion of constraints on $\sigma_\chi$, the cross section of SHDM with protons or very low $A$ nuclei, from astrophysics and cosmology. In astrophysics, some constraints come from limits on heating. One example is the constraint 
from limits on cooling rates of H I regions in the interstellar medium. The limits are \cite{Chivukula:1989cc}
\begin{equation}
    \sigma_\chi\ls 10^{-16} \ {\rm cm^2}\Biggl(\frac{m_\chi}{10^7\ {\rm GeV}}
    \Biggr) \frac{1}{f_\chi}\,,
\end{equation}
using the average of the energy loss per nucleon in H I regions, 
$\lambda=(8.1\pm 4.8)\times 10^{-14}$ eV/s. More stringent limits on $\sigma_\chi$ have been set based galactic halo infall times, so as not to disrupt normal matter disk of the Milky Way \cite{Starkman:1990nj,Natarajan:2002cw}, on the order of $\sigma_\chi\ls 5\times 10^{-18} \ {\rm cm^2}\,({m_\chi}/{10^7\ {\rm GeV}})$, however, these limits assume $f_\chi=1$.

DM heating of cold gas clouds near the Galactic center provide additional constraints \cite{Bhoonah:2018gjb}. 
In Ref. \cite{Bhoonah:2018gjb}, limits on $\sigma_{\chi N}$ can be set by requiring the heating from DM scattering be less than the average volumetric cooling rate of cold gas clouds near the Galactic Center. The limits assumed $A^4$ scaling, so even with an iron mass fraction in the cloud of $f_{\rm Fe}=0.0014$, DM scattering with iron dominates over DM scattering with hydrogen by a factor of $4\times 10^4$. For reference, depending on the cloud, cloud model and metallicity, $\sigma_{\chi N}$ is constrained to be approximately
\begin{equation}
    \sigma_{\chi N}\ls (2\times 10^{-25}-2\times 10^{-22})\ {\rm cm^2} \Biggl(\frac{m_\chi}{10^7\ {\rm GeV}}\Biggr)\frac{1}{f_\chi}\,.
\end{equation}
It is assumed that the DM loses at most half of its kinetic energy in transit to the center of the cold gas cloud, a distance
of order $R\sim 10$ pc. This means that for $\sigma_{\chi N}$ larger than $10^{-18}-4\times 10^{-16}$ cm$^2$, the bounds are not valid.

Following the same constraints, one can determine the corresponding limits on $\sigma_c\simeq \sigma_\chi$ with the volumetric cooling limit. The $\sigma_\chi$ cross section is approximately $\sigma_c$ for these gas clouds because $f_H=0.71$, in contrast to the Earth which is more than 1/3 iron by mass. The range of upper bounds from cooling limits is 
\begin{equation}
    \sigma_\chi\simeq \sigma_c\ls (3\times 10^{-21}-6\times 10^{-18})\ {\rm cm^2} \Biggl(\frac{m_\chi}{10^7\ {\rm GeV}}\Biggr)\frac{1}{f_\chi}\,.
\end{equation}
The approximations break down for $\sigma_c
\gs (6\times 10^{-14}-5\times 10^{-13})$ cm$^2$.
The most stringent limits come from the coldest cloud modeled,  labeled as ``preliminary'' because of on-going characterization of the average cloud temperature.
This means that the interactions and energy loss occur on the outer edges of the cold gas clouds where the limits are not applicable. It is noted in ref. \cite{Bhoonah:2018gjb} that the gas clouds have hot exteriors, with volumetric cooling rates as much as two to three orders of magnitude larger than the average for the cloud. Dedicated modeling of DM heating of the clouds for large cross section may be required \cite{Bhoonah:2018gjb}.

Constraints on dark matter cross sections with nuclei from Big Bang Nucleosynthesis come primarily from DM inelastic, rather than elastic, scattering with deuterium. The constraint on the DM deuterium break-up cross section is very weak \cite{Cyburt:2002uw}, in any case. Constraints from the flux of gamma rays from cosmic ray interactions with DM \cite{Cyburt:2002uw} are not clearly applicable here, as the necessary interaction is one that produces $\pi^0$'s that decay to photons, not the elastic cross section we discuss here.

Cosmological limits are the most constraining of the DM-baryon cross section $\sigma_\chi$.
Cosmological constraints are based on modeling density perturbations in the early Universe with the inclusion of DM-baryon interactions. Limits from data from the Planck satellite and Sloan Digital Sky Survey yield \cite{Dvorkin:2013cea}
\begin{equation}
    \sigma_\chi\ls 5\times 10^{-20}\ {\rm cm^2}\,\Biggl(\frac{m_\chi}{10^7\ {\rm GeV}}\Biggr)\,,
\end{equation}
assuming all of the dark matter is super heavy.
Recently, a new evaluation of constraints from modeling the cosmic microwave background and baryon acoustic oscillations (CMB+BAO) yield \cite{Buen-Abad:2021mvc}
\begin{eqnarray}
\sigma_\chi&\ls &8\times 10^{-22}\ {\rm cm^2}\Biggl(\frac{m_\chi}{10^7\ {\rm GeV}}\Biggr){\ \rm for}\ f_\chi=1\\
\sigma_\chi&\ls &2\times 10^{-19}\ {\rm cm^2}\Biggl(\frac{m_\chi}{10^7\ {\rm GeV}}\Biggr){\ \rm for}\ f_\chi=0.01\,.
\end{eqnarray}
These CMB+BAO constraints are among the most stringent on DM-baryon cross sections.

A DM-baryon cross section of $\sigma_\chi\sim 10^{-20}$ cm$^2$ would permit efficient capture of SHDM by the Earth such that $C_\chi^\oplus$ is independent of cross section. With $f_\chi\sim 0.01-0.1$, such a large cross section is acceptable from the astrophysical and cosmological points of view.
A sufficiently long Earth drift time for SHDM to allow for a DM distribution throughout the Earth is more difficult to achieve.
The breakdown of the $A^4$ scaling relation between $\sigma_{\chi N}$ and $\sigma_{\chi A}$ make it difficult to conclusively exclude or admit $\sigma_\chi\sim 10^{-20}$ cm$^2$,
absent a specific DM composite model.
Furthermore, given the very slow drift velocities, a more complete treatment of Earth dynamics including convection in the mantle may be required.
 
\bibliographystyle{utphys}
%\bibliography{../references}
\providecommand{\href}[2]{#2}\begingroup\raggedright\endgroup

\end{document}